\documentclass[aps,preprint,prd,showpacs,nofootinbib]{revtex4}

\usepackage{amsmath,amssymb,amsfonts,array}
\usepackage{bm,bbm}
\usepackage{cases}
\usepackage{dcolumn}
\usepackage{graphicx}
\usepackage{hyperref}
\usepackage{subfigure}
\usepackage{wasysym}
\usepackage{xcolor}

\definecolor{dullred}{rgb}{0.706,0.208,0.192}
\definecolor{darkred}{rgb}{0.545,0,0}
\definecolor{MaroonC}{rgb}{0,0.502,0.502}
\definecolor{dullblue}{rgb}{0,0.298,0.49}
\definecolor{blue3}{RGB}{31, 119, 180}
\definecolor{dullpurple}{rgb}{0.431,0.188,0.534}
\definecolor{darkgreen}{rgb}{0.075,0.302,0.047}
\definecolor{darkergreen}{rgb}{0,0.196,0.125}
\definecolor{darkergreen2}{rgb}{0,0.294,0.188}

\hypersetup{colorlinks=true,
	breaklinks=true,
	pdfstartview=Fit,
	linkcolor=blue,
	citecolor=blue,
	urlcolor=blue}

\def\be{\begin{equation}}
\def\ee{\end{equation}}
\def\ba{\begin{eqnarray}}
\def\ea{\end{eqnarray}}

\def\lf{\left}
\def\rt{\right}

\begin{document}
	

\title{Generating enhanced parity-violating gravitational waves during inflation with violation of the null energy condition}

\author{Yong Cai}

\email{yongcai\_phy@outlook.com}

\affiliation{School of Physics and Microelectronics, Zhengzhou University, Zhengzhou, Henan 450001, China}

\begin{abstract}

A violation of the null energy condition (NEC) during inflation in a single-field inflation model will naturally enhance the amplitude of the parity violation effect (defined by $\Delta\chi$) of inflationary primordial gravitational waves (GWs), provided the inflaton is non-minimally coupled to a gravitational Chern-Simons term. After going through the NEC-violating phase, the Universe enters subsequent slow-roll inflation with a higher energy scale (i.e., a greater Hubble parameter $H$), which results in an enhanced nearly scale-invariant power spectrum (i.e., $P_{\rm T}$) of inflationary primordial GWs in the high-frequency band, while $P_{\rm T}$ remains consistent with observations in the frequency band of the cosmic microwave background.
Therefore, the violation of NEC during inflation will amplify the observability (i.e., $P_{\rm T}\cdot\Delta\chi$) of the parity violation effect on small scales. Intriguingly, our model has particular oscillatory features on $\Delta\chi$ that may not be mimicked by others.


\end{abstract}

\maketitle
\tableofcontents

\section{Introduction}

The inflation scenario \cite{Guth:1980zm,Linde:1981mu,Albrecht:1982wi,Starobinsky:1980te} has achieved tremendous success in simultaneously solving several puzzles of the hot big bang cosmology. The nearly scale-invariant power spectrum of the primordial scalar perturbations predicted by inflation has been confirmed by observations of the temperature anisotropy of cosmic microwave background (CMB) \cite{Planck:2018vyg,Planck:2018jri}, while the primordial gravitational waves (GWs) predicted by inflation remains undetermined.
For a scale-invariant power spectrum of the primordial GWs, the tensor-to-scalar ratio is bounded to be $r_{0.002} < 0.035$ at $95\%$ confidence level \cite{BICEP:2021xfz} in the observational window of CMB. In recent years, the detections of GWs from binary black holes \cite{LIGOScientific:2016aoc} and a
binary neutron star inspiral \cite{LIGOScientific:2017vwq} by the LIGO and Virgo collaborations have opened a new window for gravitational physics.

The inflationary primordial GWs background \cite{Starobinsky:1979ty,Rubakov:1982df}, which spans a broad frequency band (about $10^{-18}-10^{10}$ Hz), may bring us a wealth of information about gravity and the early Universe in light of the recent and future observations in a wide multi-frequency range, including those of the Pulsar Timing Array (PTA) \cite{NANOGrav:2020bcs,Goncharov:2021oub,Hobbs:2009yy,Kramer:2013kea}, SKA \cite{Weltman:2018zrl}, LISA \cite{LISA:2017pwj}, Taiji \cite{Ruan:2018tsw}, TianQin \cite{TianQin:2020hid}, DECIGO \cite{Kawamura:2011zz} and BBO \cite{Crowder:2005nr}.
In contrast to primordial scalar perturbations, primordial GWs possess a distinct characteristic, namely, chirality, which can be manifested in parity-violating theories of gravity in the primordial Universe, see e.g. \cite{Lue:1998mq,Balaji:2003sw,Alexander:2004wk,Satoh:2007gn,Satoh:2008ck,Li:2009rt,Cai:2016ihp,Smith:2016jqs,Gubitosi:2016yoq,Obata:2016oym,Bartolo:2017szm,Yagi:2017zhb,Yoshida:2017cjl,Bartolo:2018elp,Nishizawa:2018srh,Gao:2019liu,Nojiri:2019nar,Mylova:2019jrj,Fujita:2020iyx,Nojiri:2020pqr,Chu:2020iil,Li:2020xjt,Fu:2020tlw,Cai:2021uup,Li:2022mti,Li:2022vtn,Martinovic:2021hzy,Kamada:2021kxi,Zhang:2022xmm,Odintsov:2022hxu,Peng:2022ttg,Bastero-Gil:2022fme,Jiang:2022uxp,Sulantay:2022sag,Qiao:2022mln} for some of the recent studies, see also \cite{Alexander:2004us,Maleknejad:2011sq,Maleknejad:2011jw,Kawai:2017kqt,Odintsov:2019mlf,Bombacigno:2022lcx,Odintsov:2021kup,Boudet:2022nub,Bombacigno:2022naf,Odintsov:2022cbm}.
A dynamical coupling of the inflaton or a spectator field to a parity-violating term, such as the gravitational Chern-Simons (gCS) term \cite{Jackiw:2003pm,Alexander:2009tp}, which is well motivated \cite{Green:1984sg}, is able to produce asymmetric right- and left-handed circularly polarised isotropic GWs, see also \cite{Cai:2016ihp} for the chirality oscillation.
Notably, a hint of parity-violating physics in the CMB (i.e., the cosmic birefringence) was recently extracted from the polarization data of Planck and WMAP \cite{Minami:2020odp,Diego-Palazuelos:2022dsq,Eskilt:2022cff}, which may require further verification.
This signal of birefringence (if confirmed by more evidence) must have been generated by some mechanism beyond the slow-roll inflation, see, e.g., \cite{Fujita:2022qlk}.

In single-field inflationary models, the effect of parity violation is proportional to $\dot{\phi}\equiv {\rm d}\phi/{\rm d} t$ (where $\phi$ is the inflaton), which is suppressed by the slow-roll condition ($\dot{\phi}\ll H$), unless a delicate design of the coupling is introduced. Furthermore, the stringent bound on the tensor-to-scalar ratio in the observational window of CMB makes the situation even worse. Consequently, the parity violation of primordial GWs generated in single-field models of slow-roll inflation can hardly be observed in general situations.
However, the approximate $60$ $e-$folds inflationary expanding history could be more complicated than that described by the slow-roll inflation, see e.g., \cite{Ramirez:2011kk,Liu:2013kea,Cai:2015nya,Mishima:2019vlh,Cai:2020qpu,Bernal:2020ywq,Kuroyanagi:2020sfw,Giare:2020vss,DAmico:2020euu,DAmico:2021fhz,Lewicki:2021xku,Giare:2022wxq}. As can be inferred, the scale invariance of the GW power spectrum could be broken in such a broad frequency band through many different ways  \cite{Kuroyanagi:2014nba,Cai:2015dta,Cai:2015ipa,Cai:2015yza,Cai:2016ldn,Li:2016awk,Cai:2020ovp,Liu:2020mru,Benetti:2021uea}.
Recently, the NANOGrav Collaboration reported evidence for a stochastic common-spectrum process
\cite{NANOGrav:2020bcs}, which might be interpreted as a stochastic GW background with a spectrum tilt $-1.5\lesssim n_T\lesssim 0.5$.
If this result could be attributed to the primordial GWs, it would mean that we may need new physics beyond the slow-roll inflation, see e.g., \cite{Vagnozzi:2020gtf}.

One possibility is when there is a violation of the null energy condition (NEC) during inflation, which might play
a significant role in the evolution history of the Universe, see \cite{Rubakov:2014jja} for a review.
Implementing a fully stable NEC violation in cosmology is a challenging task \cite{Rubakov:2014jja,Libanov:2016kfc,Kobayashi:2016xpl,Ijjas:2016vtq,Dobre:2017pnt}, see also \cite{Dubovsky:2005xd,Creminelli:2006xe,Nicolis:2009qm,Rubakov:2013kaa,Elder:2013gya,Battarra:2014tga,Koehn:2015vvy,Qiu:2015nha,Cai:2022ori}. It has be demonstrated with the effective field theory method that a fully stable NEC violation can be realized in ``beyond Horndeski'' theory \cite{Cai:2016thi,Creminelli:2016zwa,Cai:2017tku,Cai:2017dyi,Kolevatov:2017voe}, see also \cite{Cai:2017dxl,Cai:2017pga,Mironov:2018oec,Qiu:2018nle,Ye:2019sth,Ye:2019frg,Mironov:2019qjt,Akama:2019qeh,Tahara:2020fmn,Ilyas:2020qja,Ilyas:2020zcb,Zhu:2021whu,Zhu:2021ggm,Ageeva:2021yik,Mironov:2022ffa,Mironov:2022quk,Wolf:2022yvd} for recent studies. The violation of NEC is able to enhance the power spectrum of primordial GWs in the higher frequency band \cite{Cai:2020qpu,Cai:2022nqv}, where the scale invariance is recovered due to the recovery of slow-roll condition in the subsequent inflationary phase, which has a larger Hubble parameter $H$.
Furthermore, the slow-roll condition will be inevitably violated around the NEC-violating phase, which is able to naturally enhance the amplitude of the parity violation effect of inflationary primordial GWs in single-field inflationary models, as we will show in the following.
To our knowledge, the enhanced parity-violating GWs produced by a NEC-violating phase during inflation has not been
investigated so far.

In this paper, we illustrate the ability of a NEC violation in generating enhanced parity-violating primordial GWs during inflation with a (non-canonical) single-field model of inflation, in which the inflaton is non-minimally coupled to the gCS term.
We show that the violation of NEC during inflation will naturally amplify the observability of the parity violation effect, which might be detectable at scales much smaller than the CMB scale in the future.

\section{NEC violation and Parity violation of inflationary gravitational waves}\label{sec:2212-1}

The gCS term is the leading-order parity-violating correction to general relativity motivated by the
anomaly cancellation in particle physics and string theory, see e.g. \cite{Alexander:2009tp} for a review.
In this section, we consider a non-minimal coupling of the inflaton to the gCS term, while setting the background evolution to the same as that introduced in \cite{Cai:2020qpu}.

Initially, the canonical scalar field $\phi$ (i.e., the inflaton) slowly rolls down a nearly flat potential $V(\phi)$, so that the Universe undergoes a period of slow-roll inflation, which is NEC-preserving. The primordial scalar perturbation and GWs generated in this period should be consistent with observations in the window of the CMB. As the field $\phi$ becomes non-canonical and climbs up the potential rapidly such that ${\dot H}>0$, the NEC and the slow-roll condition are violated. After exiting the NEC-violating phase, the Universe gradually enters a subsequent slow-roll inflationary phase (NEC-preserving) with higher energy scale (i.e., a greater Hubble parameter), during which the inflaton $\phi$ becomes canonical again.

\subsection{Setup}

The action of our single-field model can be given as follows,
\be\label{genesis-action}
S=\int d^4x\sqrt{-g}\Big[{M_p^2\over 2}R - M_p^2 {g_1(\phi)\over 2}X + {g_2(\phi) \over 4}X^2 -M_p^4 V(\phi)  + {g_3(\phi)\over 8} R\wedge R +L_{\rm HD}   \Big] \,,
\ee
where the inflaton $\phi$ is dimensionless, $X=\nabla_{\mu}\phi\nabla^\mu\phi$, $g_n(\phi)$ are dimensionless functions; the gCS term $R\wedge R=\epsilon^{\alpha\beta\gamma\delta}
R_{\alpha\beta\mu\nu}R_{\gamma\delta}^{~~\mu\nu}$, $\epsilon^{\alpha\beta\gamma\delta}$ is the four dimensional Levi-Civita tensor with $\epsilon^{0123}=-1/\sqrt{-g}$; $L_{\rm HD}$ represents some higher derivative terms which are assumed to have negligible contributions to the background evolution and the tensor perturbations at quadratic order, see e.g. \cite{Cai:2017dyi} for $L_{\delta g^{00}R^{(3)}}$. In this paper, we will focus on the primordial tensor perturbations.
The dangerous ghost or gradient instabilities (see, e.g., \cite{Battarra:2014tga,Koehn:2015vvy,Libanov:2016kfc,Kobayashi:2016xpl}) of the scalar perturbations around the NEC-violating phase are assumed to be cured by $L_{\rm HD}$ (see, e.g., \cite{Cai:2016thi,Creminelli:2016zwa,Cai:2017tku,Cai:2017dyi,Kolevatov:2017voe}, for $L_{\rm HD}=L_{\delta g^{00}R^{(3)}}$), for simplicity.
A detailed study of the scalar sector of this theory will be carried out separately.


With the flat Friedmann-Robertson-Walker metric,  i.e., $ds^2=-dt^2+a^2(t)d\vec{x}^2$, the background equations can be found as
\ba \label{eqH} &\,& 3 H^2
M_p^2=\frac{M_p^2}{2} g_1\dot{\phi }^2 +\frac{3}{4} g_2 \dot{\phi
}^4 +M_p^4 V \,,
\\&\,&
\dot{H} M_p^2= -\frac{M_p^2}{2} g_1 \dot{\phi }^2 -\frac{1}{2} g_2
\dot{\phi }^4\,,  \label{dotH}
\ea
where
``$_{,\phi}\equiv d/d\phi$''.
The background evolution is unaffected by the gCS term or $L_{\rm HD}$.
By choosing the functions $g_{1}(\phi)$, $g_{2}(\phi)$ and $V(\phi)$ appropriately, we can realize the background evolution of intermittent NEC violation during inflation, see \cite{Cai:2020qpu}.

In the unitary gauge, we will set the tensor perturbation as $h_{i j}=a^{2} \left(\mathrm{e}^{\gamma}\right)_{i j}$, where $\gamma_{i i}=0=\partial_{i} \gamma_{i j}$. The quadratic action of tensor perturbation can be given as
\ba
S_{\gamma}^{(2)}={ M_{p}^2\over8 } \int d\tau d^3x\lf\{a^2\Big[
{\gamma_{ij}'}^2-(\partial_k \gamma_{ij})^2 \Big]-{g_3'\over M_p^2
}\epsilon^{ijk}\lf[(\partial_i\gamma_{jl})'(\gamma_{k}^{l})'
-\partial_i\partial_l\gamma_{jq}\partial^l\gamma_{k}^{q} \rt]\rt\}\,,\label{Lhh221017}
\ea
where $'=d/d\tau$ and $\tau=\int dt/a $, $\epsilon^{ijk}=\epsilon^{0ijk}$ is the three dimensional Levi-Cevita symbol, the higher derivative terms proportional to $g_3'$ are the parity-violating corrections contributed by the gCS term.
During the slow-roll inflation, the effect of parity violation could be suppressed by $\dot{\phi}$ since $g_3'=a g_{3,\phi}\dot{\phi}$, unless a delicate design of $g_3(\phi)$ is adopted.
However, the parity violation during inflation could be greatly enhanced by an intermediate NEC violation, during which $\dot{\phi}$ can naturally and significantly violate the slow-roll condition.

In the Fourier space, we have $\gamma_{ij}(\tau,\mathbf{x})=\sum_{s={\rm L},{\rm R}}\int{d^3\mathbf{k}\over
(2\pi)^3}\gamma^{(s)}_{\mathbf{k}}(\tau)p_{ij}^{(s)}(\mathbf{k})e^{i\mathbf{k}\cdot
\mathbf{x}}$,
where $p_{ij}^{(s)}$ is the circular polarization tensor which
satisfies $p_{ij}^{({\rm R})}{p^{ij}}^{({\rm R})}=p_{ij}^{({\rm L})}{p^{ij}}^{({\rm L})}=0$,
$p_{ij}^{({\rm R})}{p^{ij}}^{({\rm L})}=2$ and
$ik_l\epsilon^{nlj}{p_{ij}}^{(s)}=k\lambda^{(s)}{p^n_i}^{(s)}$.
The parameters $\lambda^{({\rm L})}=-1$ and $\lambda^{({\rm R})}=1$ correspond
to the left- and the right-handed modes, respectively.
For convenience, we also define $\lambda^{({\rm N})}=0$, which corresponds to the situation where there is no violation of parity.

From Eq. (\ref{Lhh221017}), we can obtain the equation of motion of $\gamma_{\mathbf{k}}^{(s)}$ as
\be
{u^{(s)}_{\mathbf{k}}}''+\lf[ \lf(c_{{\rm T} \mathbf{k}}^{(s)}\rt)^2 k^2-{
{z_{\rm T}^{(s)}}''\over z_{\rm T}^{(s)} }\rt]u^{(s)}_{\mathbf{k}}=0\,,
\label{eomu1}
\ee
where we have defined ${u^{(s)}_{\mathbf{k}}}\equiv z_{\rm T}^{(s)}{\gamma_{\mathbf{k}}^{(s)}}$ and
$\quad z_{\rm T}^{(s)}={a\over2}\sqrt{1-\lambda^{(s)}{k\over
a^2}{g_3'\over M_p^2 } }$, and the effective sound speed is $c_{{\rm T} \mathbf{k}}^{(s)}=1$ for both the left- and the right-handed GW modes, see e.g., \cite{Nishizawa:2018srh}.
In order to avoid the appearance of the ghost modes, we require that ${k\over
a^2}{g_3'\over M_p^2 }<1$, see e.g., \cite{Dyda:2012rj}.

The perturbation modes are deep inside their horizon initially, i.e.,
$k^2\gg {z_{\rm T}^{(s)}}''/{z_{\rm T}^{(s)}}$, which indicates ${u^{(s)}_{\mathbf{k}}}\simeq
e^{-i k\tau}{/\sqrt{2 k}}$ for $\tau\rightarrow-\infty$. The power
spectrum of primordial GWs is $P_{\rm T}^{(s)}={k^3\over2\pi^2}\lf|\gamma_{\mathbf{k}}^{(s)}\rt|^2$ for $k\ll aH$, where $s=$ L, R and N.\footnote{Note that $P_{\rm T}^{(N)}=P_{\rm T}/2$ in the situation where there is no parity violation.}
To evaluate the intensity of parity violation of primordial GWs,
it is convenient to define the chiral parameter
\be \Delta\chi={ P_{\rm T}^{({\rm L})}-P_{\rm T}^{({\rm R})} \over P_{\rm T}^{({\rm L})}+P_{\rm T}^{({\rm R})} } \,.\label{eq:Deltachi01}
\ee
From an observational point of view, what we really care about is the difference between the power spectra of the left- and right-handed GW modes, i.e., $P_{\rm T}^{({\rm L})}-P_{\rm T}^{({\rm R})}$.
With a sufficient large $P_{\rm T}= P_{\rm T}^{({\rm L})}+P_{\rm T}^{({\rm R})}$, a $\Delta\chi$ as small as a few percent in the GW background might be detectable for future observations. Furthermore, the peculiar features on $\Delta\chi$ may encode the characteristic of our model.


\subsection{Model and numerical solution}

We adopt the phenomenological model proposed in \cite{Cai:2020qpu}, in which we set
\ba
g_1(\phi)&=&
{2\over 1+e^{-q_1(\phi-\phi_0)}}+
{1\over 1+e^{q_2(\phi-\phi_3)}} -{f_1 e^{2\phi} \over 1+f_1
e^{2\phi} }\,,\label{g1}
\\
g_2(\phi)&=& {f_2\over 1+e^{-q_2(\phi-\phi_3)}} {1\over
    1+e^{q_3(\phi-\phi_0)}}\,,
\quad
g_3(\phi)= \phi\,,
\label{g3}
\\ V(\phi)&=&{1\over2}m^2\phi^2 {1\over
    1+e^{q_2(\phi-\phi_2)}}+
{\lambda\over
    1+e^{-q_4(\phi-\phi_1)}} \left[1-\frac{(\phi-\phi_1)^{2}}{\sigma^{2}}\right]^{2}\,, \label{V}
\ea
where $m$, $\lambda$, $f_{1,2}$ and $q_{1,2,3,4}$ are some positive
constants. We set $\phi_3<\phi_2<0<\phi_1<\phi_0$ for convenience.

In the limit $\phi\ll \phi_3$, we have $g_1= 1$, $g_2= 0$ and
$V=V_{\rm inf1}\simeq m^2\phi^2/2$, which is responsible for the first stage of slow-roll inflation. Apparently, the inflaton is canonical when $\phi\ll \phi_3$.
For $\phi\gg \phi_0$, we
have $g_1= 1$, $g_2= 0$ and $V=V_{\rm inf2}\simeq \lambda
[1-\frac{(\phi-\phi_1)^{2}}{\sigma^{2}}]^{2}\gg
V_{\rm inf1}$, which is responsible for the subsequent slow-roll inflation with a higher energy scale (or equivalently, a greater Hubble parameter $H$). Again, the inflaton becomes canonical in this stage.
In these two NEC-preserving regimes (i.e., $\phi\ll \phi_3$ and $\phi\gg \phi_0$), the potential is nearly flat, so that the predicted primordial scalar perturbations are consistent with the observations in the window of CMB.

\begin{figure}[htbp]
    \subfigure[~~$\dot{\phi}$ and $H$]{\includegraphics[width=0.46\textwidth]{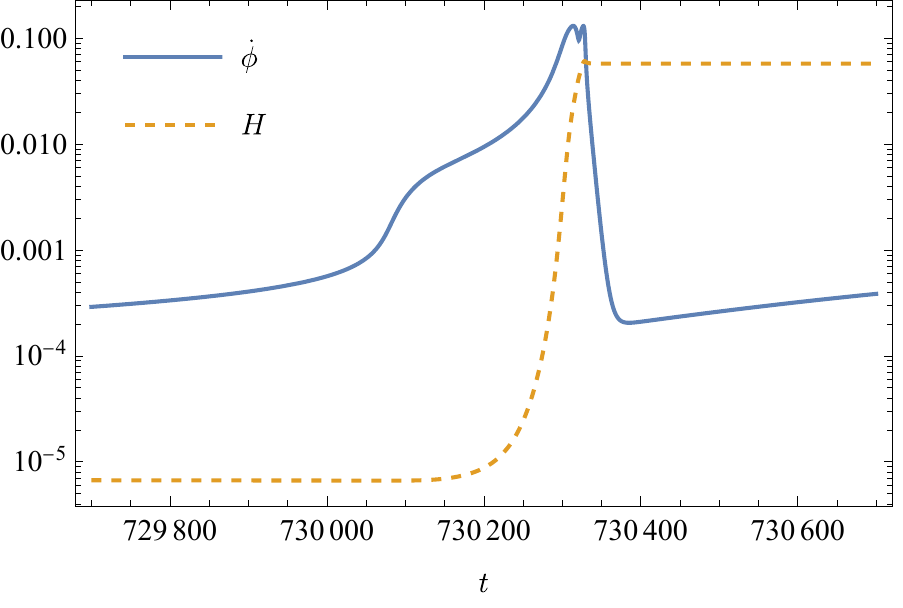} }
    \subfigure[~~$\ddot{\phi}$]{\includegraphics[width=.50\textwidth]{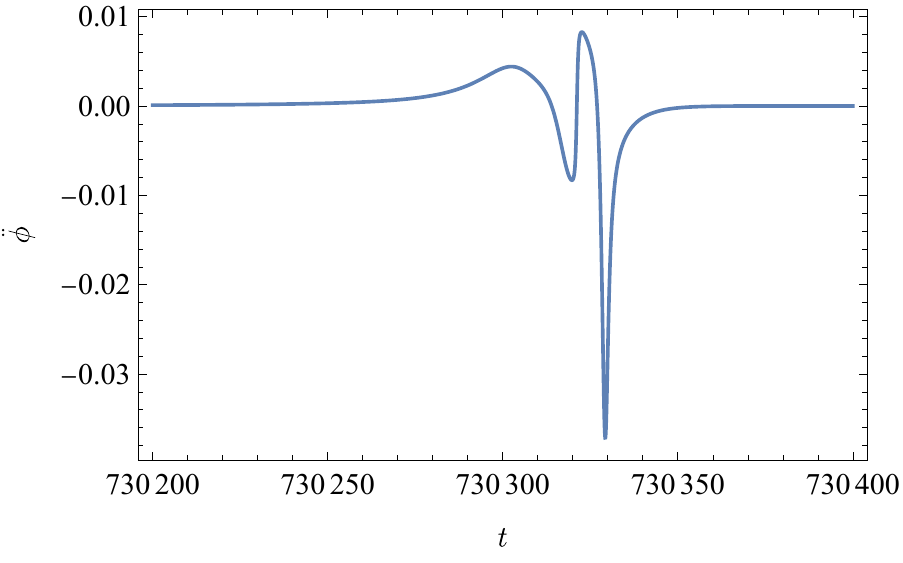} }
    \caption{Evolution of the background. Approximately, $|\dot{\phi}|$ and $|\ddot{\phi}|$ reach their maximum values around the end of the NEC violation phase.} \label{NECC-figDDotphi-1}
\end{figure}

\begin{figure}[htbp]
\centering %
    \includegraphics[width=0.55\textwidth]{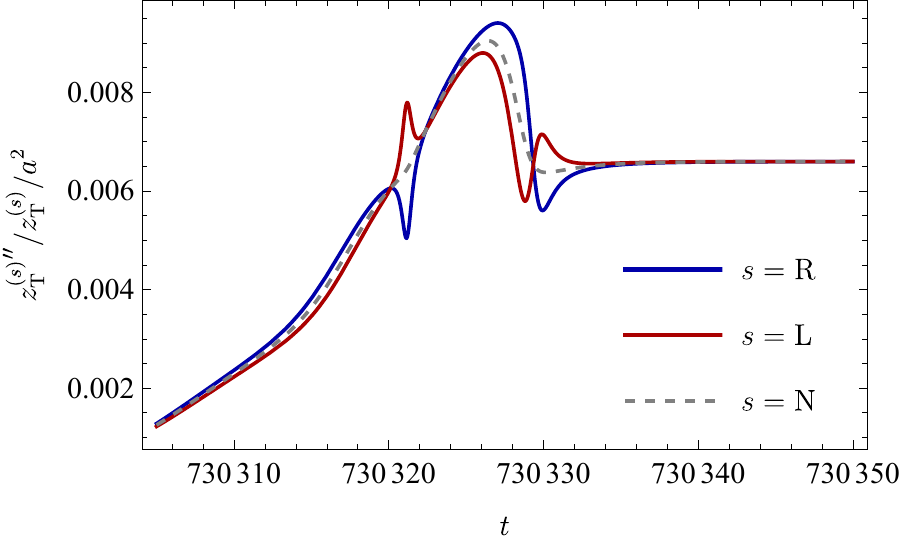}
    \caption{Comparison of the evolutions of ${z_\text{T}^{(s)}}''/{z_\text{T}^{(s)}}/a^2$ for $s={\rm R}$ ($\lambda_s=1$), $s={\rm L}$ ($\lambda_s=-1$) and $s={\rm N}$ ($\lambda_s=0$), where $s={\rm N}$ corresponds to $a''/a^3$. We have set $k=8.8\times 10^5$ $\text{Mpc}^{-1}$.} \label{NECC-figzppbz-1}
\end{figure}

We numerically solve Eqs. (\ref{eqH}) and (\ref{dotH}) using (\ref{g1}) to (\ref{V}), and show the evolutions of $H$, $\dot{\phi}$ and $\ddot{\phi}$ around the NEC-violating phase in Fig. \ref{NECC-figDDotphi-1}. Apparently, the slow-roll condition is strongly violated around the NEC-violating phase.
More details about the background evolution can be found in \cite{Cai:2020qpu}.
The evolution of ${z_\text{T}^{(s)}}''/{z_\text{T}^{(s)}}$ is displayed in Fig. \ref{NECC-figzppbz-1}.
Actually, it is mainly the nontrivial evolutions of $\dot{\phi}$ and $\ddot{\phi}$ that cause ${z_{\rm T}^{(s)}}''/{z_{\rm T}^{(s)}}$ to deviate greatly from $a''/a$, where $a''/a$ corresponds to the situation in which parity is preserved (i.e., $\lambda^{(s)}=0$). Therefore, the NEC violation during inflation is able to naturally enhance the parity violation, compared with the case where the slow-roll condition is preserved, as can inferred.

\begin{figure}[htbp]
    \subfigure[~~$|u|_\mathbf {k}^{(s)}$]{\includegraphics[width=0.47\textwidth]{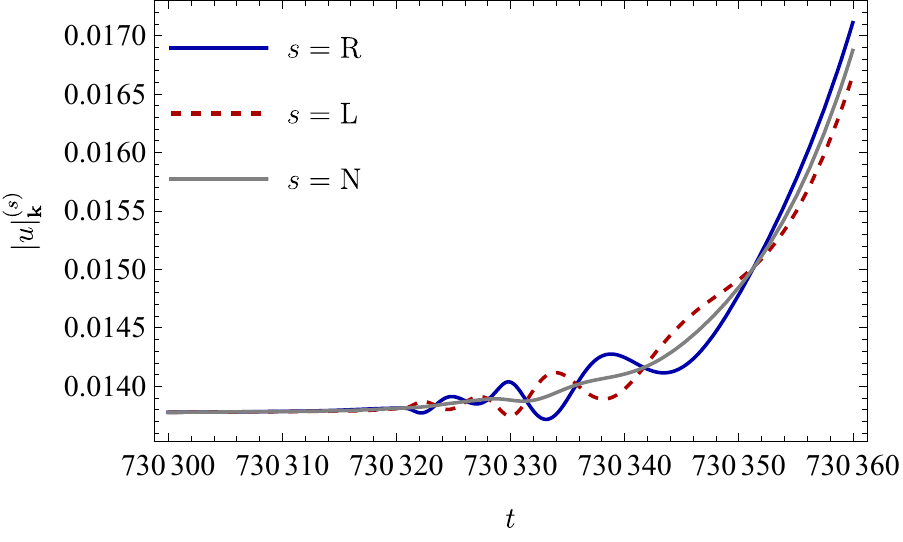} }
    \subfigure[~~$|u|_\mathbf {k}^{{(s)}}/|u|_\mathbf{k}^{\text{(N)}}$]{\includegraphics[width=.47\textwidth]{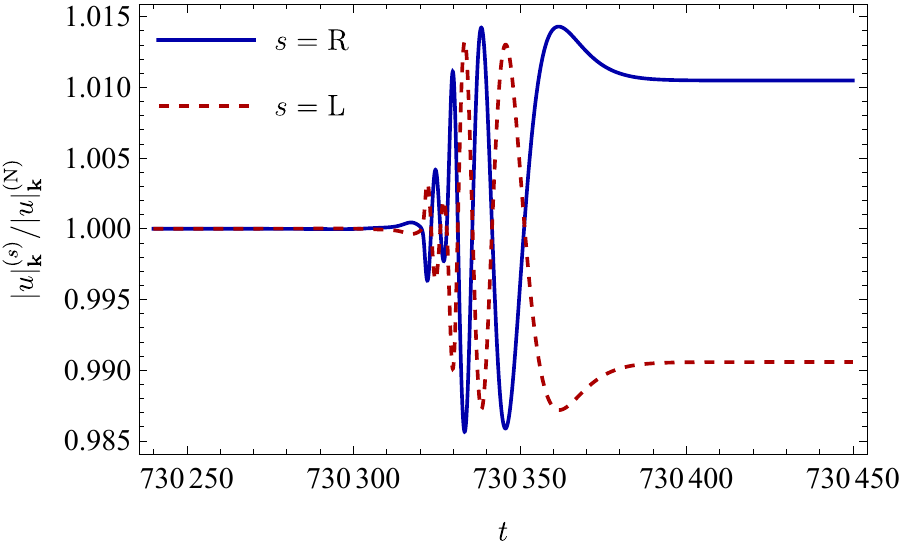} }
    \subfigure[~~$|\gamma|_\mathbf {k}^{(s)}$]{\includegraphics[width=0.46\textwidth]{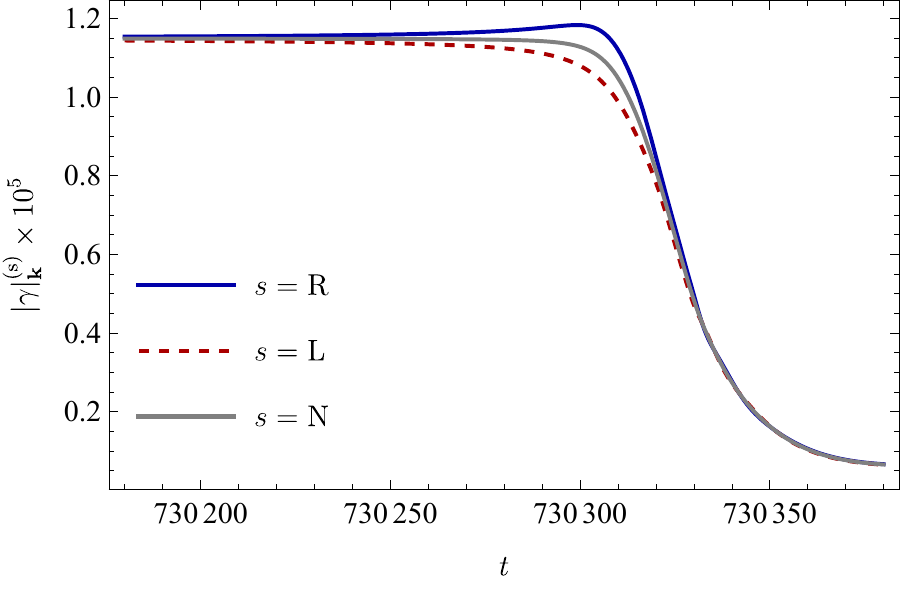} }
    \subfigure[~~$|\gamma|_\mathbf{k}^{(s)}/|\gamma|_\mathbf{k}^{\text{(N)}}$]{\includegraphics[width=.47\textwidth]{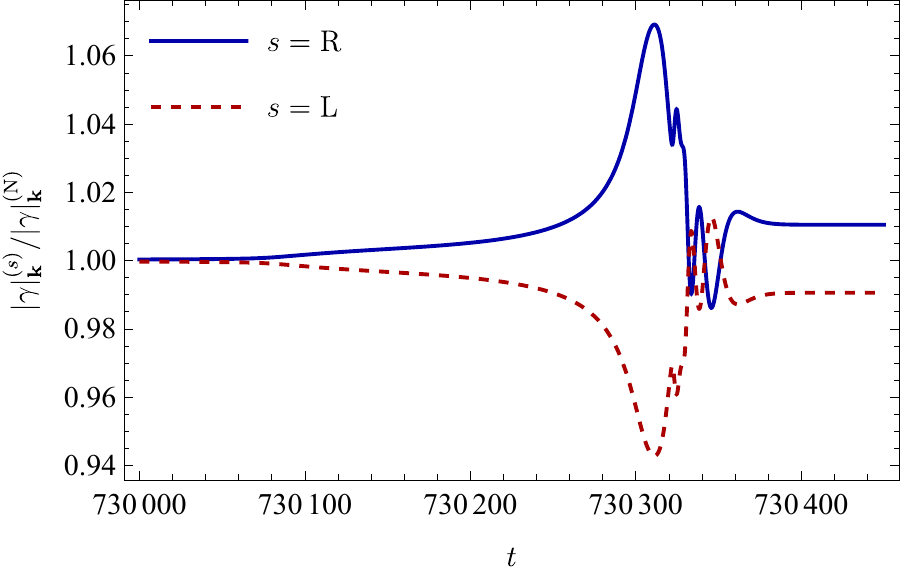} }
    \caption{Evolution of the GW mode with comoving wavenumber $k=8.8\times 10^6$ $\text{Mpc}^{-1}$.} \label{NECC-Fig-u-2}
\end{figure}

We also numerically solve Eq. (\ref{eomu1}) and plot the evolutions of perturbation modes $|u|_\mathbf {k}^{(s)}$ and $|\gamma|_\mathbf {k}^{(s)}$ for $s=$L, R and N in Fig. \ref{NECC-Fig-u-2}, in which we set the comoving wavenumber $k=8.8\times 10^6$ $\text{Mpc}^{-1}$.
Basically, the chirality asymmetry of GW modes becomes relatively apparent when $|\dot{\phi}|$ and $|\ddot{\phi}|$ become sufficient large, as we can see from Figs. \ref{NECC-figDDotphi-1} to \ref{NECC-Fig-u-2}.  The left- and right-handed GW modes oscillate with each other and eventually arrive at different values after exiting their horizon. As a result, the parity violation of primordial GWs enhanced by NEC violation during inflation will be encoded in the power spectra, see also \cite{Cai:2016ihp} for the chirality oscillation.

\begin{figure}[htbp]
    \subfigure[~~$P_{\rm T}^{({\rm s})}/P_{\rm T,inf2}^{({\rm s})}$]{\includegraphics[width=0.49\textwidth]{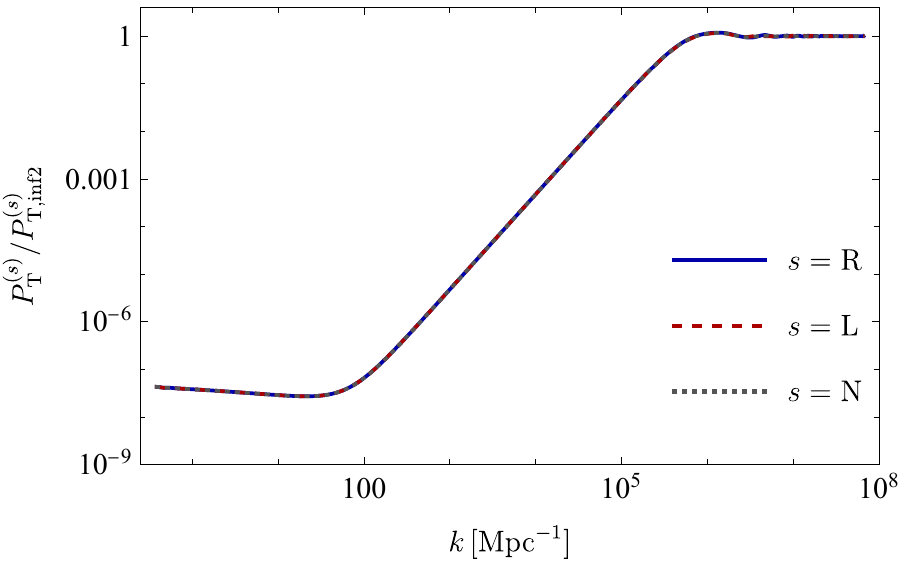} }
    \subfigure[~~$P_{\rm T}^{({\rm s})}/P_{\rm T,inf2}^{({\rm s})}$]{\includegraphics[width=.47\textwidth]{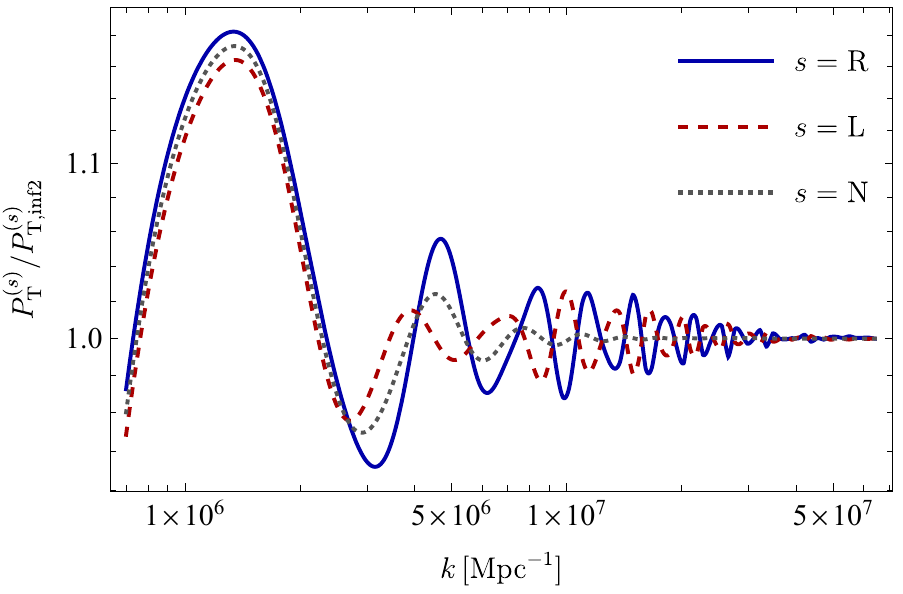} }
    \subfigure[~~$P_{\rm T}^{(s)}/P_{\rm T}^{({\rm N})}$]{\includegraphics[width=0.47\textwidth]{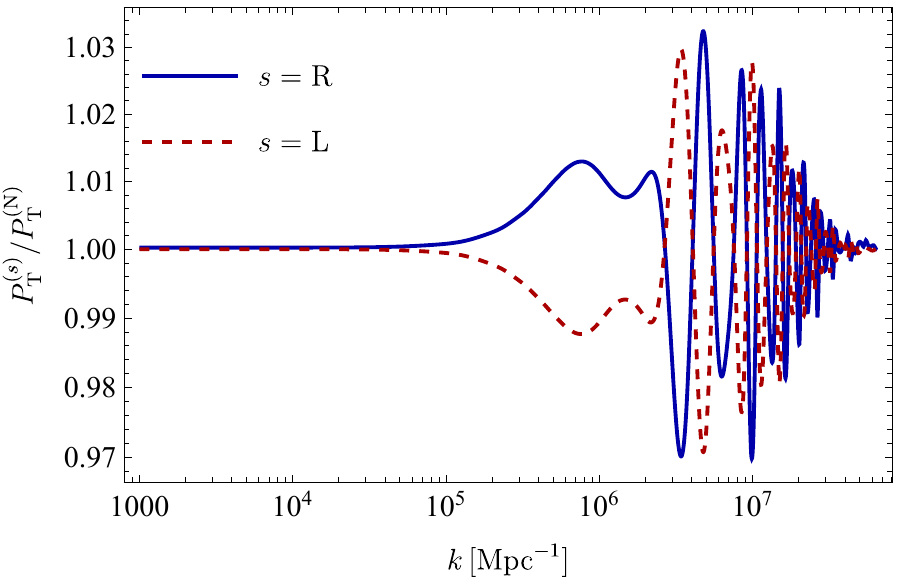} }
    \subfigure[~~$\lf(P_{\rm T}^{({\rm L})}+P_{\rm T}^{({\rm R})}\rt)/P_{\rm T}^{({\rm N})}/2$]{\includegraphics[width=.49\textwidth]{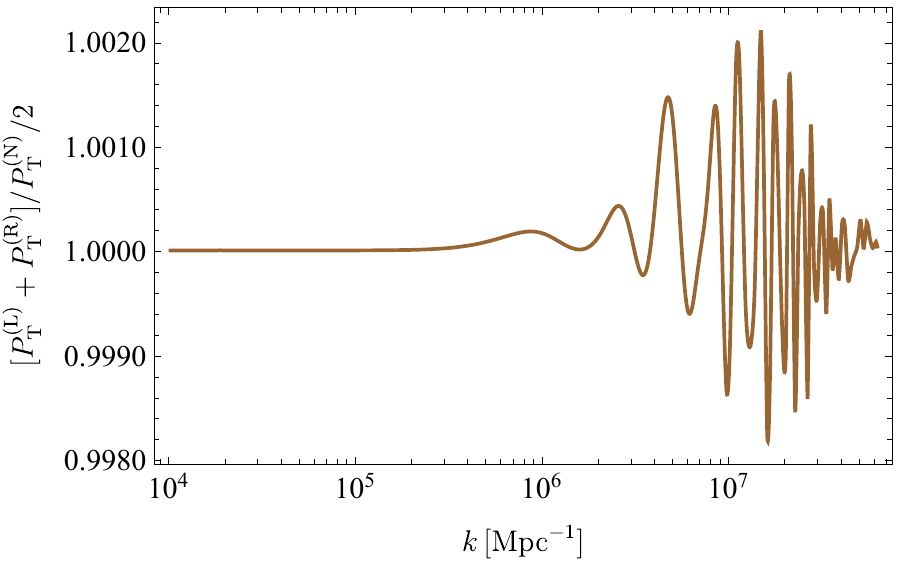} }
    \caption{Comparisons of the power spectra of the left-handed ($\lambda^{(s)}=-1$, $s={\rm L}$), right-handed ($\lambda^{(s)}=1$, $s={\rm R}$) and parity preserving ($\lambda^{(s)}=0$, $s={\rm N}$) primordial GWs modes which exited their horizon around the NEC-violating phase (part a) and that around the end of the NEC-violating phase (part b). The ratios of the power spectra of the chiral GW modes to that of the parity preserving ones (parts c and d). Here, the power spectra are nearly scale-invariant on both large and small scales, which may be distinctive from that produced by other mechanism.} \label{NECC-Fig-1}
\end{figure}

\begin{figure}[htbp]
\centering
    \includegraphics[width=0.47\textwidth]{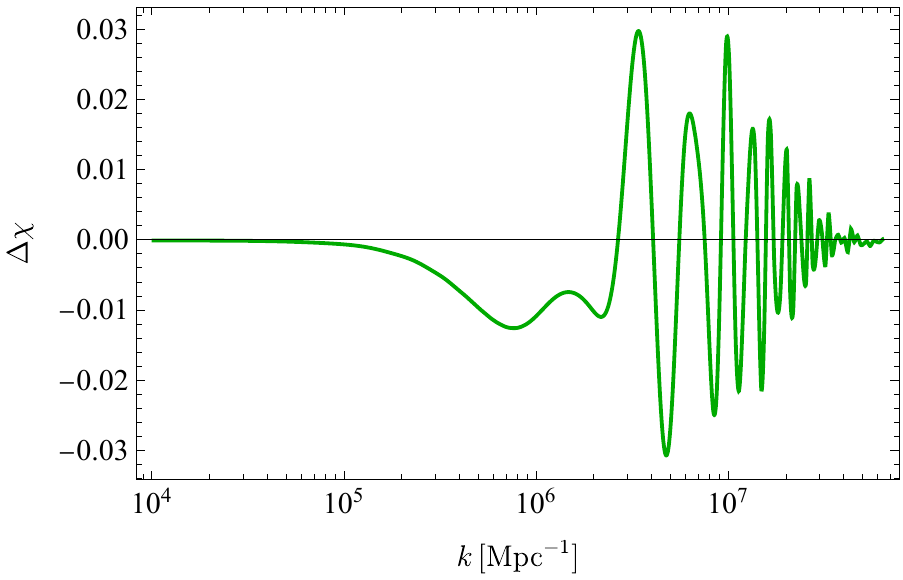}
    \caption{The chiral parameter $\Delta\chi$ is defined by Eq. (\ref{eq:Deltachi01}). These nontrivial oscillations of $\Delta\chi$ become evident in the range $10^5 {\rm Mpc}^{-1} \lesssim k \lesssim 10^8 {\rm Mpc}^{-1}$, which corresponds to those GW modes exiting horizon around the end of the NEC-violating phase.} \label{NECC-Fig-3}
\end{figure}

The power spectra and the chiral parameter $\Delta\chi$ are displayed in Figs. \ref{NECC-Fig-1} and \ref{NECC-Fig-3}, respectively.
We can see that the parity violation becomes evident in the range $10^5 {\rm Mpc}^{-1} \lesssim k \lesssim 10^8 {\rm Mpc}^{-1}$. This range corresponds to those GW modes exiting horizon around the end of the NEC-violating phase, around which $|\dot{\phi}|$ and $|\ddot{\phi}|$ reach their maximum values. It should be pointed out that the specific relation between the comoving wavenumber $k$ and the time $t$ depends on the choice of parameters in the numerical calculation. For example, with a different normalization of the scale factor, we can obtain a shift of the power spectra as well as $\Delta\chi$ with respect to $k$.

Notably, we can see in Fig. \ref{NECC-Fig-3} that there are some peculiar oscillatory features on $\Delta\chi$ at
small scales. These features together with the peculiar power spectra (as shown in Fig. \ref{NECC-Fig-1}) make our model characteristic with respect to others in the literature (e.g., \cite{Obata:2016oym,Cai:2021uup,Zhang:2022xmm,Peng:2022ttg,Bastero-Gil:2022fme}).

%

\section{Parity-violating gravitational wave background and observations}\label{sec:2212-2}

The energy density spectrum of GWs is defined as \cite{Turner:1993vb,Boyle:2005se} (see also \cite{Kuroyanagi:2014nba})
\begin{equation}
\label{density} \Omega_{\rm GW} = \frac{1}{\rho_{\text{c}}}\frac{d\rho_{\rm GW}}{d\ln
k}=\frac{k^{2}}{12 a^2_0H^2_0}P_{\rm T}{\cal T}^2(k,\tau_0)\,,
\end{equation}
where the critical energy density of the Universe is $\rho_{\text{c}}=3M_p^2 H^{2}_0$, $H_0=67.8\, {\rm km/s/Mpc}$, $\tau_{0}=1.41\times10^{4}$ Mpc, $a_0=1$, the reduced Hubble parameter $h=H_0/(100\,{\rm km/s/Mpc})$, $\rho_{\text{GW}}$ is
the energy density of GWs at present, and the wavenumber relates to the frequency as $k=2\pi f$.
The transfer function can be given as
\be
{\cal T}(k,\tau_{0})=\frac{3
\Omega_{\text{m}}j_1(k\tau_0)}{k\tau_{0}}\sqrt{1.0+1.36\frac{k}{k_{\text{eq}}}+2.50\lf(\frac{k}{k_{\text{eq}}}\rt)^{2}},
\label{Tk}
\ee
where $k_{\text{eq}}=0.073\,\Omega_{\text{m}} h^{2}$
Mpc$^{-1}$ is the comoving wavenumber of the perturbation mode that entered the
horizon at the equality of matter and radiation, $\Omega_{\rm m}$ is the density fraction of matter today.

Currently, the bound on the tensor to scalar ratio put by data of CMB \cite{BICEP:2021xfz} indicates that $\Omega_{\rm GW}<10^{-15}$ at $f\sim 10^{-16}$ Hz.
The 12.5-yr data of NANOGrav \cite{NANOGrav:2020bcs} suggests $\Omega_{\rm GW}\sim 10^{-9}$ at $f\sim 10^{-8}$ Hz, provided the stochastic common-spectrum process can be interpreted as a stochastic GW background.
The observations in the frequency band of LIGO and Virgo \cite{LIGOScientific:2016aoc,LIGOScientific:2017vwq} put the bound $\Omega_{\rm GW}<10^{-7}$ at $f\simeq 30$ Hz.

\begin{figure}[htbp]
    \subfigure[~~]{\includegraphics[width=0.48\textwidth]{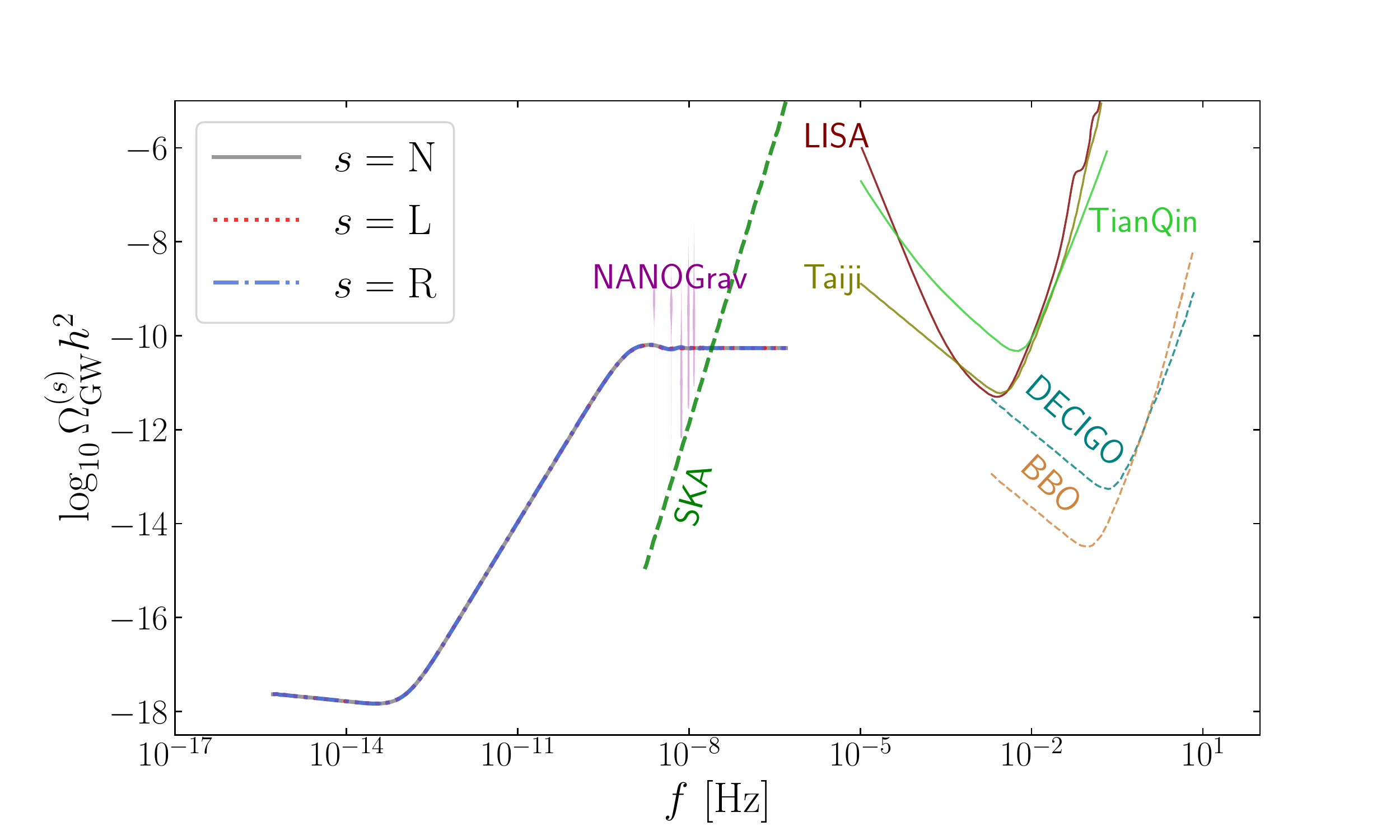} }
    \subfigure[~~]{\includegraphics[width=0.48\textwidth]{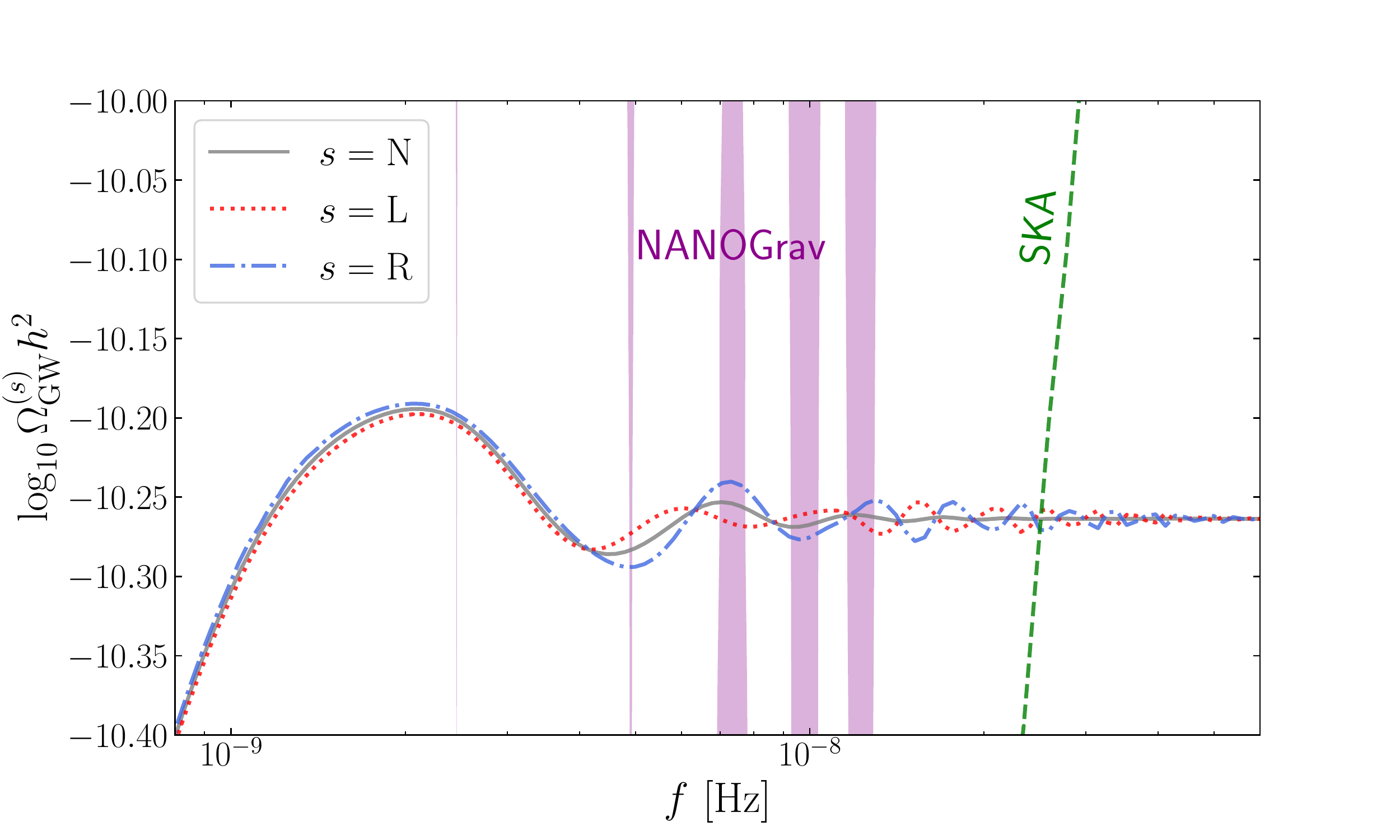} }
    \caption{The energy density spectrum of GWs $\Omega_{\rm GW}^{(s)}h^2$ with respect to the physical frequency $f$, where
$h=H_0/(100{\rm km/s/Mpc})$; $s=$L and R correspond to the left- and right-handed GW modes, respectively; $s=$N corresponds to the parity-preserving mode. The largest parity violation effect appears in the band of PTA.} \label{Fig-221208-1}
\end{figure}

\begin{figure}[htbp]
    \subfigure[~~]{\includegraphics[width=0.48\textwidth]{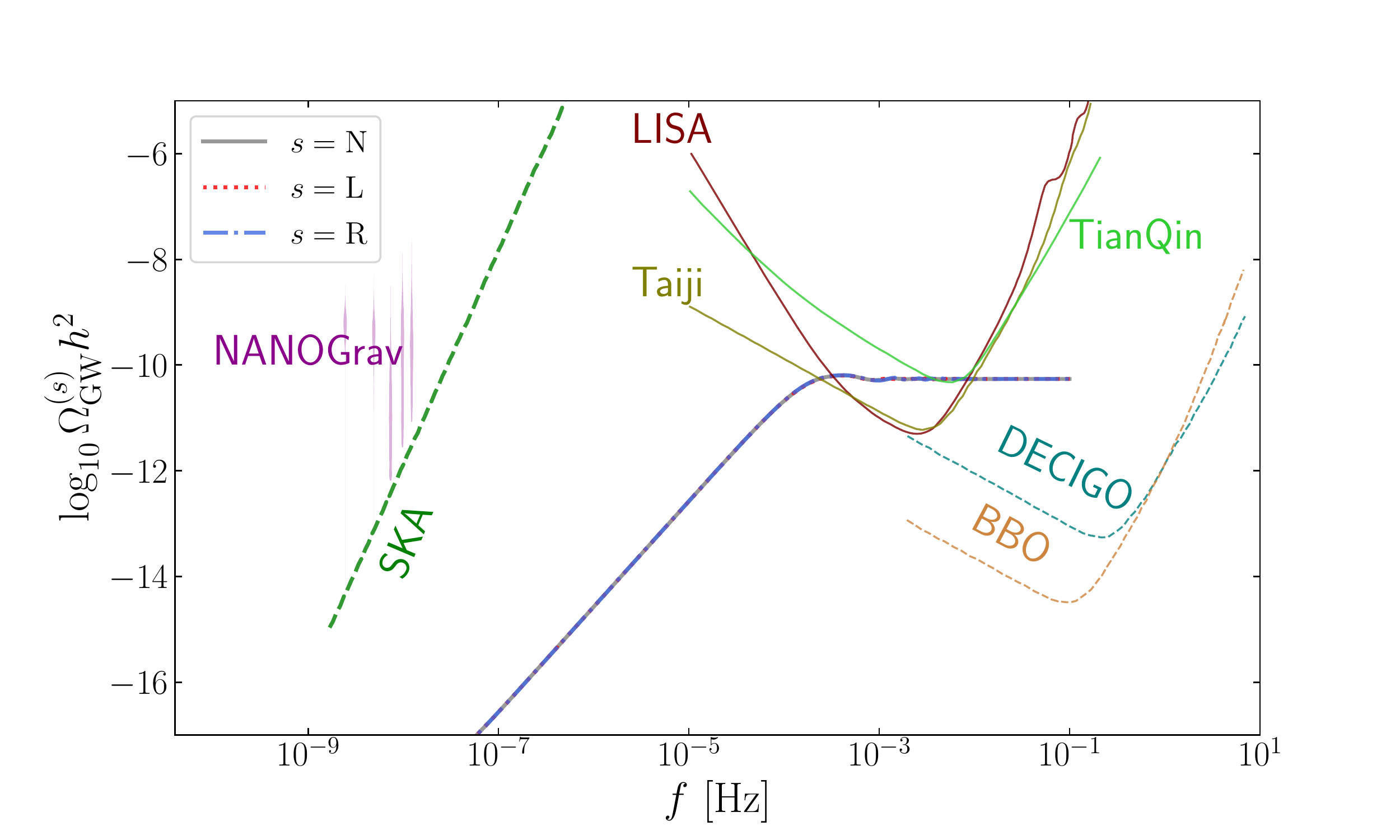} }
    \subfigure[~~]{\includegraphics[width=0.48\textwidth]{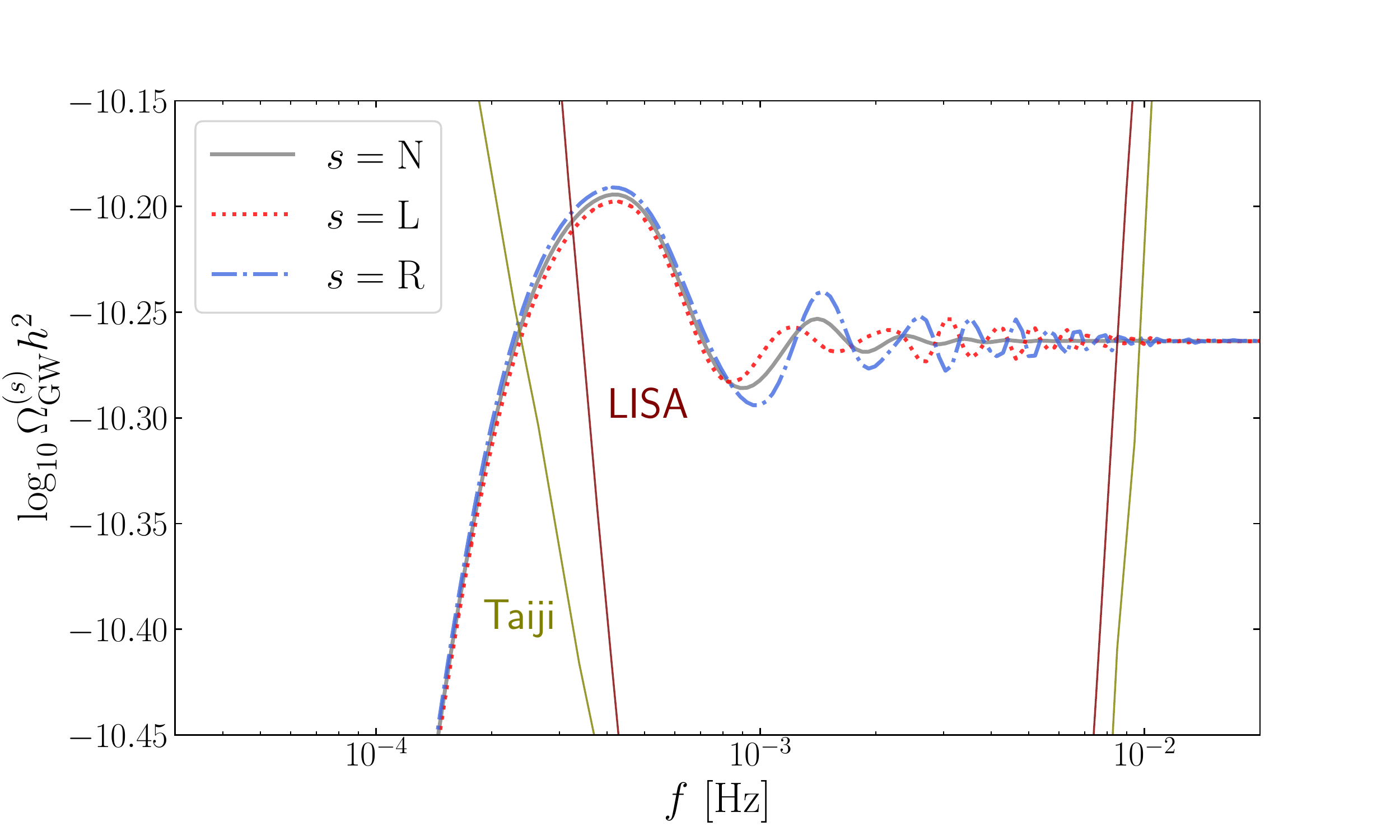} }
    \caption{The energy density spectrum of GWs $\Omega_{\rm GW}^{(s)}h^2$ with respect to the physical frequency $f$, where
$h=H_0/(100{\rm km/s/Mpc})$; $s=$L and R correspond to the left- and right-handed GW modes, respectively; $s=$N corresponds to the parity-preserving mode. The largest parity violation effect appears in the band of space-based interferometers.} \label{Fig-221208-2}
\end{figure}

We plot $\Omega_{\rm GW}^{(s)}h^2\sim P_{\rm T}^{(s)}$ with respect to the physical frequency $f$ for left-, right-handed and parity-preserving GW modes in Fig. \ref{Fig-221208-1}. We use the 12.5-yr data of NANOGrav and the future sensitivity curves of SKA, LISA, Taiji, TianQin, DECIGO and BBO to compare the theoretical predictions with experimental projections. In Fig. \ref{Fig-221208-1}, the largest parity violation effect appears in the observational window of PTA.
We also choose a different normalization of the scale factor in our numerical calculation so that we can shift the largest parity violation effect to the band of space-based interferometers, as displayed in Fig \ref{Fig-221208-2}.

Additionally, with a different choice of the parameter space, we should obtain a larger effect of parity violation, i.e., a larger maximum value of $|\Delta\chi|$. Here, for simplicity and to show that the NEC violation during inflation can naturally enhance the effect of parity violation in primordial GWs, we have chosen the same model parameters in the numerical calculation as in \cite{Cai:2020qpu}.
Furthermore, since the GW power spectra $P_{\rm T}$ (or $\Omega_{\rm GW}h^2$) are greatly enhanced for more than seven orders by the NEC violation, the absolute value of $\lf|P_{\rm T}^{({\rm L})}-P_{\rm T}^{({\rm R})}\rt|$ (or $\Omega_{\rm GW}^{({\rm L})}h^2-\Omega_{\rm GW}^{({\rm R})}h^2$) could become quite substantial, even though the maximum value of $|\Delta\chi|$ is only a few percent.
Therefore, the NEC violation is able to amplify the observability of parity violation in the GW background.
It is also interesting to note that observations of an enhanced $P_{\rm T}$ and the peculiar oscillatory features on $\Delta\chi$ might distinguish our model from others in the literature (e.g., \cite{Obata:2016oym,Cai:2021uup,Zhang:2022xmm,Peng:2022ttg,Bastero-Gil:2022fme}).

We did not extend the curve to higher frequencies in Figs. \ref{Fig-221208-1} and \ref{Fig-221208-2} because there is a cutoff to avoid the appearance of ghost modes \cite{Dyda:2012rj} in gCS theory, as can be seen from the expression of $z_\text{T}^{(s)}$. It is also because we did not deal with the exit of the second inflationary phase, which is an open question. Interestingly, there might be multistage inflation with intermittent NEC violations and short periods of decelerated expansion, as has been discussed in \cite{Cai:2020qpu}. As a result, the GW spectra might look like the Great Wall while satisfying constraints in the band of CMB.

\section{Conclusion}

Chirality might be an intrinsic characteristic of GWs.
We have shown that a violation of the NEC during inflation is able to naturally enhance the parity-violating
primordial GWs such that the chirality asymmetry of GWs can be observable at scales much smaller than the CMB scale. We have illustrated this new mechanism with a single-field inflationary model, in which the inflaton $\phi$ is non-minimally coupled to a gravitational Chern-Simons term.

The enhancement of parity-violating GWs in our model is twofold. First, because of the strong violation of the slow-roll condition induced by the violation of NEC, $\Delta\chi$ is naturally enhanced compared with that in the case where the slow-roll condition is preserved, though $\Delta\chi$ cannot approach $|\Delta \chi| \sim 1$ in the current model. Second, the power spectrum of parity-violating GWs (i.e., $P_{\rm T}$) can be enhanced for more than seven orders in higher frequency band by the NEC violation during inflation, which might be detectable at small scales in the future, while the power spectrum remains consistent with constrains in the observational window of CMB.
Therefore, the violation of NEC will amplify the observability of the parity violation effect in the GW background, since $P_\text{T}\cdot\Delta\chi$ is the difference between the power spectrum of the left- and right-handed GWs.


An enhanced nearly scale-invariant power spectrum of the parity-violating inflationary primordial GWs in the high-frequency band with peculiar oscillatory features on $\Delta\chi$ could be distinguishable from that generated by other mechanism (e.g. \cite{Obata:2016oym,Cai:2021uup,Zhang:2022xmm,Peng:2022ttg,Bastero-Gil:2022fme}). A wide range of multi-frequency observations aimed at searching for GW signals will bring us rich information about the gravity and the early Universe. A signal of parity-violating GW background found by future observations could be attributed to that generated by our mechanism.

\acknowledgments

We would like to thank Yun-Song Piao, Gen Ye, Hao-Hao Li, Mian Zhu and Chao Chen for helpful discussions.
Y. C. is supported in part by the National Natural Science Foundation of China (Grant No. 11905224), the China Postdoctoral Science Foundation (Grant No. 2021M692942) and Zhengzhou University (Grant No. 32340282).
We acknowledge the use of the computing server {\it Arena317}@ZZU.


\bibliography{ref}
\bibliographystyle{utphys}

\end{document}